\documentclass[11pt]{article}

\usepackage{hyperref}

\usepackage{euscript}

\usepackage{latexsym}
\usepackage{amsmath, amssymb,graphics}
\usepackage[all]{xy}
\usepackage{color}

\usepackage{tikz}
\input xy
\xyoption{all}
\usepackage{euscript}
\usepackage{multirow}
\usepackage{etex, pictexwd,dcpic}
\usetikzlibrary{positioning}
\usetikzlibrary{shapes.geometric}
\usetikzlibrary{shapes.misc}
\usetikzlibrary{calc}
\usetikzlibrary{positioning}

\usepackage{pgflibraryarrows}
\usepackage{pgflibrarysnakes}

\usetikzlibrary{trees} 
\usetikzlibrary[trees] 

\usepackage{fullpage}

\newtheorem{theorem}{Theorem}[section]
\newtheorem{definition}[theorem]{Definition}
\newtheorem{lemma}[theorem]{Lemma}

\newtheorem{proposition}[theorem]{Proposition}

\numberwithin{equation}{section}

\renewcommand{\(}{\begin{equation*}}
\renewcommand{\)}{\end{equation*}}
\newcommand{\bea}{\begin{eqnarray*}}
\newcommand{\eea}{\end{eqnarray*}}
\newcommand{\R}{{\mathbb R}}
\newcommand{\C}{{\mathbb C}}
\newcommand{\Z}{{\mathbb Z}}

\def\proof {{Proof.}\hspace{7pt}}

\def\endofproof {\hfill{$\Box$}\\}

\newcommand{\beq}{\begin{equation}}
\newcommand{\eeq}{\end{equation}}

\numberwithin{equation}{section}

\renewcommand{\(}{\begin{equation}}
\renewcommand{\)}{\end{equation}}

\def\R{{\mathbb R}}
\def\Z{{\mathbb Z}}

\def\C{{\mathbb C}}

\def\1{{\bf 1}}

\def\<{\langle}
\def\>{\rangle}

\numberwithin{equation}{section}

\renewcommand{\(}{\begin{equation}}
\renewcommand{\)}{\end{equation}}

\begin{document}

\title{Ninebrane Structures}

 \author{
  Hisham Sati\\
  }

\maketitle

\begin{abstract} 
String structures in degree four are associated with cancellation of anomalies of string theory
in ten dimensions. Fivebrane structures in degree eight have recently been shown to be 
associated with cancellation of anomalies associated to fivebranes in 
string theory and M-theory. We introduce and describe 
 {\it Ninebrane structures}  in degree twelve and demonstrate how they 
capture some anomaly cancellation phenomena in M-theory. 
Along the way we also define certain variants, considered as intermediate cases in
degree nine and ten, 
which we call {\it 2-Orientation} and
{\it 2-Spin structures}, respectively.  
As in the lower degree cases,
we also discuss the natural twists of these structures and characterize the corresponding 
topological groups  associated to each of the structures, which likewise admit 
refinements to differential cohomology.

\end{abstract}


\tableofcontents

\newpage
\section{Introduction}

The study of higher connected covers of Lie groups in the context of string 
theory and M-theory, as advocated in \cite{SSS1, SSS2, SSS3}, leads to 
interesting mathematical structures as well as means for canceling anomalies 
in string theory and M-theory. Beyond String structures in degree four, obtained by 
killing 
the third homotopy group of the orthogonal group, 
we have Fivebrane structures   in degree eight obtained by killing the next homotopy group which is in
degree seven. 
 
%

\vspace{2mm}
We will consider killing -- more precisely, co-killing -- 
 further homotopy groups. From the homotopy theoretic point of view
one can continue the process of killing indefinitely in a systematic way. However, 
no systematic understanding of the relevance of all cases exists. What we do is advocate 
is a natural setting, a description of the higher geometry, as well as provide several examples
from M-theory  and string theory for which performing such killings in the next few degrees 
is natural.  We highlight the structures we consider here in the following table. 

{\small
  \begin{center}
\hspace{-.5cm}
 \begin{tabular}{|c||cccccccccl}
 \hline
   $k$ & $7$ & $8$ & $9$ & $10$ & $11$ & $12$ & &
   \\
   \hline
   \hline
   &&&&&&&&&&
   \\
$\begin{array}{cc} {\rm Homotopy~groups}\\ 
  \pi_k(\mathrm{O}(n))
  \end{array}$
 &  $\Z$ & $\mathbb{Z}_2$ & $\Z_2$ &   $0$ & $\Z$ & $0$ & &
   \\
   \hline
   &&&&&&&&&
   \\
   $\begin{array}{cc}
   {\rm Connected~ covers} \\ \mathrm{O}(n)\langle k\rangle
   \end{array}$
   &
   $\mathrm{String}(n)$
   &
   $\mathrm{Fivebrane}(n)$
   &
   \fbox{$\mathrm{O\langle 9\rangle}(n)$}
   &
   \fbox{$\mathrm{O\langle 10\rangle}(n)$}
   &
   &
   \fbox{$\mathrm{Ninebrane}(n)$}
   &
   &
   \\
   &
   \multicolumn{9}{l}{
     \xymatrix@C=4pt{
        \hspace{30pt}
        \ar@{|->}@/_1pc/[rr]_{\mathrm{kill}\; \pi_7}&&
        \hspace{50pt}
        \ar@{|->}@/_1pc/[rr]_{\mathrm{kill}\; \pi_8}&&
        \hspace{30pt}
        \ar@{|->}@/_1pc/[rrr]_{\mathrm{kill}\; \pi_9}&&&&
        \hspace{-10pt}
        \ar@{|->}@/_1pc/[rrrrrrr]_{\mathrm{kill}\; \pi_{11}}
        &&&&&&&&&&&
     }}\\
\hline
 \end{tabular}
 \end{center}
}

\medskip
The  point of view we take here is that the 
group $O\langle 9 \rangle (n)$ is a `shift by 8' analog of the special orthogonal group ${\rm SO}(n)$. 
The group $O\langle 10 \rangle (n)$ is a `shift by 8' analog of the Spin group ${\rm Spin}(n)$.  
The mod 8 periodicity of the homotopy groups of the orthogonal group motivates 
 the following for the corresponding $G$-structures: 
The classifying spaces $BO \langle 10 \rangle = B( O \langle 9 \rangle)$ and
 $BO \langle 11 \rangle = B( O \langle 10 \rangle)$ 
correspond to a `shift by 8' analog of orientation and of Spin structure, respectively.
Thus to identify these structures in the second period in the
 mod 8 periodicity we indicate these 
as {\it 2-Orientations} and {\it 2-Spin structures}. 
 We encapsulate the theme in the following diagram of lifts, extending the ones in \cite{SSS2} \cite{SSS3}, i.e. 
the higher part of the Whitehead tower of the orthogonal group:
$$
\xymatrix{
  &&&  {B}\mathrm{Ninebrane}=BO\langle 13 \rangle
    \ar[d]
	&& 
     \\
  &&&   BO\langle 11\rangle \ar[rr]^{x_{12}} 
     \ar[d]
     &&
     K(\Z, 12)
     \\
  &&&   BO\langle 10 \rangle \ar[rr]^{x_{10}}
    \ar[d]
     &&
     K(\Z_2, 10)
     \\
  &&&  {B}\mathrm{Fivebrane} \ar[rr]^{x_9} 
    \ar[d]
	&& K(\Z_2, 9)
    \\
 X 
 \ar[rrr]
 \ar[urrr]
 \ar@{..>}[uurrr]
 \ar@{..>}[uuurrr]
 \ar@{..>}[uuuurrr]
 &&&  {B} \mathrm{String} \ar[rr]^{\tfrac{1}{6}{p}_2} 
    && K(\Z, 8)\;.
     }
$$
We identify the obstructions $x_i$, $i=9,10$
in Sec. \ref{bo8} and $i=12$ in Sec. \ref{sec 9}, 
and characterize the set of lifts in Sec. \ref{sec set}. 
The first two will be, in a  sense, exotic classes.
Then in Sec. 
\ref{sec twist} we twist these structures, and in 
Sec. \ref{sec variant} we consider variant structures in which the 
higher obstructions vanish without lower classes having to be zero. 
We 
  characterize the corresponding
groups in Sec. \ref{sec group}, and finally in Sec. \ref{sec dif} we 
construct differential refinements, and provide a natural M-theoretic setting
in way of motivation and examples throughout.

\section{$BO\langle 10\rangle$ and $BO\langle 11\rangle$ structures}
\label{bo8}

The topology and geometry of a manifold can be studied via the structures related
to its tangent bundle. Starting with a Riemannian manifold $X^n$,
its tangent bundle with structure group $O(n)$, can be
lifted to further structures which in turn imposes topological
conditions on $X^{n}$. The first step is to lift the structure group
to $SO(n)$ by equipping $X^n$ with an orientation, which is allowed
provided  the first Stiefel-Whitney class $w_1$ of
$TX^n$ is zero. Further structures can be conditionally given. The structure group
can be further lifted from $SO(n)$ to the double cover ${\rm
Spin}(n)$ which allows the existence of spinors provided
that the second Stiefel-Whitney class $w_2$ is zero. 

\medskip
Note that the
process does not stop here, and we can further continue
equipping the tangent bundle with higher structures. Due to the
homotopy type of the orthogonal group, the next step in the process
is consider the lifting to the seventh connected cover denoted $O
\langle 7 \rangle$, which occurs when the cohomology class $\tfrac{1}{2}p_1$
is zero, where $p_1$ is the first Pontrjagin class of the tangent
bundle. The notation $G \langle n \rangle$ means that all the
homotopy groups of order $0, \cdots, n$ of the original group $G$
are killed.

\vspace{3mm} In fact, the way to obtain the above structures is by
pulling back from the universal classifying space to our spacetime
$X^n$. Since $SO(n)$ is obtained from $O(n)$ by killing the first
homotopy group, then $BSO \simeq BO\langle 2 \rangle$, which is also
denoted $BO\langle w_1 \rangle$ to highlight the condition imposed
by such a structure. Note that in going from $G$ to $BG$ there is a
shift in homotopy, i.e. $\pi_i (G)\cong \pi_{i+1}(BG)$. Similarly, $B{\rm
Spin} \simeq BO\langle 4 \rangle \simeq BO\langle w_1, w_2 \rangle$,
this time emphasizing the Spin condition $w_1=0=w_2$. Finally,
$BO\langle 8 \rangle$, sometimes also denoted $B{\rm String}$, can
be written in the same notation as $BO \langle w_1, w_2,
\frac{1}{2}p_1 \rangle$. Notice that in this last case, the
additional condition is no longer a mod 2 condition but rather is one on 
integral cohomology. The requirement $\frac{1}{2}p_1=0$ is not quite the
same as setting $p_1=0$, because the latter is a rational condition
which misses the torsion classes in the former integral condition.

\medskip
Note that $\frac{1}{2}p_1$ is related to the Stiefel-Whitney
classes, namely its mod 2 reduction is given by $w_4$. 
We will see that, in a sense, not all Stiefel-Whitney classes are relevant, but
a special role is played by 
 the ones of the form $w_{2^j}$. For instance,  
starting from $w_1=w_2=w_4=0$
leads to $w_i=0$ for $i=1, \cdots 7$. 
Thus the only new condition
after $w_4=0$ is $w_8=0$. In terms of classifying spaces, what this
implies is that there is a lift from $BO\langle 2^r \rangle$ to
$BO\langle w_{2^j} \rangle$ for $j< r$.


\medskip
We now consider the Stiefel-Whitney classes in relatively higher degrees, $w_i$ for $2^3 \leq i < 2^4$.
For applications in even higher degrees, see \cite{OP2}. 
We start with the following observation 
for the higher Stiefel-Whitney classes as they arise in the context of M-theory. 

\begin{lemma}
Let $Y^{11}$ be an  orientable eleven-manifold. 
Then we have $w_{11}(Y^{11})=w_{10}(Y^{11})=w_9(Y^{11})=0$. 
\label{w91011}
\end{lemma}

\proof 
Let $\nu_i \in H^i(Y^{11}; \Z_2)$ be the Wu class, i.e. the unique cohomology class 
such that $Sq^i (x)= x \cdot \nu_i$ for any $x\in H^{11-i}(Y^{11};\Z_2)$. From the properties
of the Wu classes, we have $\nu_0=1$, $\nu_i=0$ for $i>5$. Wu's formula relates the 
Stiefel-Whitney classes to the Wu classes via
\(
w_k= \sum_i Sq^{k-i} \nu_i\;.
\label{Wu}
\)

\vspace{-2mm}
\noindent Using the Adem relation $Sq^i=Sq^1 Sq^{i-1}$ for $i$ odd, and that
$
Sq^1: H^{10}(Y^{11};\Z_2) \to H^{11}(Y^{11};\Z_2)
$
is trivial for orientable $Y^{11}$, gives that
$
Sq^i: H^{11-i}(Y^{11};\Z_2) \to H^{11}(Y^{11};\Z_2)
$
is zero for $i$ odd. This implies, from the definition of the Wu classes, that
$\nu_i=0$ for $i$ odd. Hence, for orientable $Y^{11}$, $\nu_i$ is zero 
unless $i$ is even and $0\leq i \leq 4$. From this and expression (\ref{Wu}) it follows that
$w_i=0$ for $i>8$. Therefore, $w_9(Y^{11})=w_{10}(Y^{11})=w_{11}(Y^{11})=0$. 
\endofproof

\vspace{-2mm}
\paragraph{Remarks. 1.} Note that the M-theory fivebrane anomaly cancelation requires
an $MO\langle 9 \rangle$ orientation, i.e. a Fivebrane structure \cite{SSS3}. 

\noindent {\bf 2.} Note that in general for orientable $Y^{11}$ with no extra structure,
 the class $w_8$ will not be zero. 

\paragraph{Example.}
Consider $Y^{11}=\C P^2 \times \C P^2 \times (S^1)^3$ or $Y^{11}=P(1,4) \times (S^1)^2$,
where $P(1,4)$ is the Dold manifold defined as follows. $P(r,s)$ is the quotient 
$(S^r \times \C P^s)/\sim$, where $(x,y) \sim (x', y')$ if and only if $x'=-x$ and 
$y'=-y$.  The Dold manifold is the total space of $\C P^s$ bundle
of complex projective spaces 
over real projective space $\R P^r$ whose total Stiefel-Whitney class is given by
\(
w(P(r,s))=(1+ e_1)^r(1+ e_1 + e_2)^{s+1}\;,
\)
where $e_1$ and $e_2$ are the generators in the cohomology groups of the
corresponding projective spaces 
$H^1(\R P^r;\Z_2)$ and 
$H^2(\C P^s;\Z_2)$, respectively. In particular, $P(1,4)$ is orientable and
has non-vanishing $w_8$. 

\medskip
We have seen above that for orientable $Y^{11}$, the class 
$w_8$ is not necessarily zero. However, integrality 
of the one-loop polynomial 
$I_8=\tfrac{1}{48}[p_2- (\tfrac{1}{2}p_1)^2]$ appearing in anomaly cancellation  
in M-theory requires that $w_8$ be in fact zero. This is because it is
the mod 2 reduction of the second Spin characteristic class.
Recall
that the integral cohomology of the classifying space of the Spin 
group is \cite{Th}
\( 
H^*(B{\rm Spin}; \Z)=\Z[Q_1,Q_2,\cdots]\oplus \gamma\;,
\) 
with $\gamma$ a 2-torsion factor, i.e. $2\gamma=0$.
The two relevant degrees are 
\bea 
H^4(B{\rm Spin};
\Z)&\cong & \Z ~~~~~~~~~~~ {\rm with~~generator}~~~ Q_1
\nonumber\\
H^8(B{\rm  Spin}; \Z)&\cong &\Z\oplus \Z ~~~~~{\rm with~~generators}~~~ Q_1^2,
Q_2\;,
\eea
where the Spin classes $Q_1$ and $Q_2$ are determined by their relation to the Pontrjagin classes
\(
p_1=2Q_1\;, \qquad \qquad p_2=Q_1^2 + 2Q_2\;.
\label{Qs}
\)
Obviously, when inverting is possible, the Spin generators are given by
$Q_1=p_1/2$ and $Q_2=\frac{1}{2}p_2 - \frac{1}{2}(p_1/2)^2$.
The mod 2 reductions of $Q_1$ and $Q_2$ are $w_4$ and $w_8$,
respectively. 
It was explained in \cite{KSpin} that 
it is useful to write 
the one-loop term in 
terms of the Spin characteristic classes
\( 
I_8=\frac{Q_2}{24}\;. 
\label{Q24} 
\)
Now $I_8$ is integral when $w_4=0$ \cite{Flux}.
The latter condition allows to define a ``Membrane structure"
\footnote{This name is introduced in Ref. \cite{tw1} 
and justified there by the fact that it arises in connection with 
anomalies associated with the membrane in M-theory.}
i.e. a structure defined by the condition $w_4=0$
\cite{tw1}.  Then $Q_2$ is certainly divisible by 2, and hence we have

\begin{lemma}
For a manifold $Y^{11}$ with a Membrane structure, we have $w_8(Y^{11})=0$.
\end{lemma}

Next, putting together the above discussion,
 we have the following observation, still motivated within the context of M-theory.

\begin{proposition}
Let $Y^{11}$ be a manifold  which admits a Fivebrane structure. 
Then all classes in $Y^{11}$ pulled back from universal classes in $H^n(BO; R)$, 
for $n \leq 11$ and $R=\Z$ or $\Z_2$, are trivial. 
\end{proposition}

\proof
This follows from statements that can be directly verified. 
Orientation requires that $w_1(Y^{11})=0$. 
A Fivebrane structure amounts to $\tfrac{1}{6}p_2(Y^{11})=0$ and requires a 
String structure, i.e. $\tfrac{1}{2}p_1(Y^{11})=0$, which in turn requires a Spin condition,
i.e. $w_2(Y^{11})=0$, as well as a Membrane condition $w_4(Y^{11})=0$. 
The Fivebrane condition further implies $w_8(Y^{11})=0$ via mod 2 reduction from the Fivebrane 
condition. 
Finally, all odd Stiefel-Whitney classes up to that degree are zero. 
This follows from the Wu formula for the action of the Steenrod algebra
  which takes the general form 
$Sq^i w_j=\sum_{t=0}^i\binom{j-i-1+t}{t}w_{i-t}w_{j+t}$ 
for $i<j$ (see \cite{St}). First, the Wu formula $Sq^1w_2=w_1w_2 + w_3$ 
gives that if $w_2=0$ then $w_3=0$. Second, the formulae $Sq^1w_4=w_1w_4 + w_5$,
$Sq^2w_4=w_2 w_4 + w_6$ and $Sq^3 w_4=w_3w_4 + w_2 w_5 + w_1 w_6 + w_7$
imply that if $w_4=0$ then the three classes $w_5$, $w_6$ and $w_7$ are zero. 
In the next degree, starting with $w_8$, the formulae 
$Sq^1w_8=w_1 w_8 + w_9$, $Sq^2 w_8=w_2w_8 + w_{10}$ and 
 $Sq^3 w_8=w_3 w_8 + w_2 w_9 + w_1 w_{10} + w_{11}$ imply that if 
 $w_8=0$ then the classes $w_9$, $w_{10}$ and $w_{11}$ are zero.  
 These last three degrees can also be deduced by appealing to the dimension 
 of the manifold $Y^{11}$, i.e. by using    
Lemma \ref{w91011}.
\vspace{-5mm}
\endofproof

\vspace{3mm}
As we saw above, these obstructions mostly vanish for dimension 
reasons in our range of dimensions. However, we will consider
bundles other than the tangent bundle; for example bundles
with structure group SO(32) rather than SO(10) or SO(11).

\paragraph{Example: Orthogonal gauge structure groups.}
Consider the orthogonal group
 $G={\rm SO}(32)$
as a structure group of a gauge bundle over our manifold. 
This is relevant in  type I and heterotic string theory in ten dimensions.
\footnote{Note that the group is more precisely 
${\rm Spin}(32)/\Z_2$, where $\Z_2$ is the 
complement of the component of the 
center which leads to SO(32). However, since we are considering 
connected covers, such differences at the level of the fundamental 
group (which is killed) will not matter for us.}
The topological role of this group in relation to global anomalies 
is highlighted in \cite{Global}, which we recast in our language
(cf. \cite{SSS1} \cite{SSS2}).  
 The degree seven generator $\pi_7({\rm SO}(32)=\Z$ is used to 
show invariance of theory on $S^3 \times S^7$ and 
to derive a quantization condition on the H-field, which can be thought of as a curvature of
a gerbe in degree three.
The  next homotopy group $\pi_9({\rm SO}(32))=\Z_2$ 
correspond to a Yang-Mills instanton on $S^{10}$ 
via the embedding ${\rm SO}(10) \hookrightarrow {\rm SO}(32)$
by viewing the Spin connection, arising from a spinor 
representation of the natural structure group SO(10) of the tangent bundle, 
as a gauge field, by viewing this in the vector representation of SO(32). 

\medskip
What about $\pi_{10}({\rm SO}(32))=\Z_2$? We interpret this in essentially
the same way as for the case of $\pi_9({\rm SO}(32)$. 
However, our setting will be M-theory in eleven dimensions rather 
than string theory in ten dimensions. We start with the eleven-dimensional 
sphere $S^{11}$ with structure group ${\rm SO}(11)$ and embed this in 
the group ${\rm SO}(32)$ and, as above, view the Spin connection 
of the former as a gauge field of the latter. 
One justification for enlarging 
of the structure group is to form a generalized connection, taking into
account the C-field terms. For instance, in \cite{DL} ${\rm SO}(32)$ is 
described as a generalized holonomy group, while in \cite{Hu} 
the group ${\rm SL}(32, \R)$ played that role; homotopically, this 
is simply the same as ${\rm SO}(32)$. 

%

\medskip
We now define the first two of the new structures and afterwards we will explain the connection to the 
above classes. 

\begin{definition} (2-Orientation structure.)
 A {\it 2-Orientation structure} is defined 
by the lift from ${\rm BO}\langle 9 \rangle={\rm BFivebrane}$ to 
${\rm BO} \langle 10 \rangle$ in the following diagram
\(
\xymatrix{
&
&
{\rm BO}\langle 10 \rangle
\ar[d]
&
&
\\
X 
\ar[rr]^-f
\ar@{..>}[urr]^{\hat{f}}
&&
{\rm BFivebrane} 
\ar[r]
&
K(\pi_9(BO), 9)
\;.
}
\)
\label{def 2or}
\end{definition}

\paragraph{Remarks}
{\bf (i)} The existence of the above fibration, as well as all the fbrations that we introduce 
below, follows from the work of Stong \cite{St} \cite{St2}. 

\noindent {\bf (ii)} Corresponding to this diagram is a class
$x_9 \in H^9({\rm BFivebrane}, \pi_9(BO))=H^9({\rm BFivebrane}, \Z_2)$. 
The map $f: X \to {\rm BFivebrane}$ lifts to 
$\hat{f}: X \to {\rm BO}\langle 10 \rangle$ 
if and only if we have the vanishing of the obstruction class 
\(
f^*x_9=0 \in H^9(X; \pi_9(BO))=H^9(X; \Z_2)\;.
\)

\noindent {\bf (iii)} One might think that we can identify $x_9=w_9$ as the cohomology ring $H^*(BO; \Z_2)$ is 
generated by the Stiefel-Whitney classes. However, as we will see shortly, this is not the case.

\noindent {\bf (iv)} $BO\langle 10 \rangle$ can also be described as a bundle 
pulled back from the path fibration  in the following diagram (as in \cite{SSS3}, which builds on \cite{St})
\(
\xymatrix{
K(\pi_9(B)), 8) 
\ar[r] 
&
{\rm BO}\langle 10 \rangle
\ar[d]
&&
PK(\pi_9(BO), 9)
\ar[d]
&
\Omega K(\pi_9(BO), 9)
\ar[l]
\\
X 
\ar[r]^f
\ar@{..>}[ur]^{\hat{f}}
&
{\rm BFivebrane} 
\ar[rr]
&&
K(\pi_9(BO), 9)\;.
&
}
\label{path1}
\)

\medskip
In the next degree we have:

\begin{definition}(2-Spin structure.)
A {\it 2-Spin structure} is defined 
by the lift  from ${\rm BO}\langle 10 \rangle$ to 
${\rm BO} \langle 11 \rangle$ in the following diagram
\(
\xymatrix{
&
&
{\rm BO}\langle 11 \rangle
\ar[d]
&
&
\\
X 
\ar[rr]^-f
\ar@{..>}[urr]^{\hat{f}}
&&
{\rm BO\langle 10 \rangle} 
\ar[r]
&
K(\pi_{10}(BO), 10)
\;.
}
\)
\label{def 2or}
\end{definition}

\paragraph{Remarks}
{\bf (i)} Corresponding to this diagram is a class
$x_{10} \in H^{10}(BO\langle 10 \rangle, \pi_{10}(BO))=H^{10}(BO \langle 10 \rangle, \Z_2)$. 
The map $f: X \to BO \langle 10 \rangle$ lifts to 
$\hat{f}: X \to {\rm BO}\langle 11 \rangle$ 
if and only if we have the vanishing of the obstruction class 
\(
f^*x_{10}=0 \in H^{10}(X; \pi_{10}(BO))=H^{10}(X; \Z_2)\;.
\)

\noindent {\bf (ii)} One might think that we can identify $x_{10}=w_{10}$ as the cohomology 
ring $H^*(BO; \Z_2)$ is generated by the Stiefel-Whitney classes. 
However, again, as we will see this is not the case.

\noindent {\bf (iii)} $BO\langle 11 \rangle$ can also be described as a bundle 
pulled back from the path fibration  in the following diagram
\(
\xymatrix{
K(\pi_{10}(B)), 9) 
\ar[r] 
&
{\rm BO}\langle 11 \rangle
\ar[d]
&&
PK(\pi_{10}(BO), 10)
\ar[d]
&
\Omega K(\pi_{10}(BO), 10)
\ar[l]
\\
X 
\ar[r]^f
\ar@{..>}[ur]^{\hat{f}}
&
{\rm BO\langle 10 \rangle} 
\ar[rr]
&&
K(\pi_{10}(BO), 10)\;.
&
}
\label{path2}
\)

\paragraph{
Identifying the generators $x_9$ and $x_{10}$.} 
As mentioned above, one might be very tempted to identify the 
obstruction classes $x_9$ and $x_{10}$ with the generators 
of $H^i(BO; \Z_2)$ for $i=9, 10$, i.e. with the Stiefel-Whitney classes
$w_9$ and $w_{10}$, respectively. However, this would be 
too simplistic and, as we will see, is not true. The main subtlety here 
is that beyond Fivebrane in the Whitehead tower of $BO$, 
maps from $BO \langle m \rangle$ 
to BO are no longer surjective (see \cite{St} \cite{St2}). 
Thus upon close inspection one realizes that
these will be exotic characteristic classes not arising 
from the cohomology of $BO$ but rather of $BO\langle 9\rangle$
and $BO \langle 10 \rangle$.  
For instance,  $x_9$ arises
from pulling back along 
the map $K(\Z, 7) \to {\rm BFivebrane}$. 
This is analogous to the map $K(\Z, 3) \to BO \langle 7 \rangle$
relating gerbes to String structures and provides interesting geometry. 
It would be very interesting to identify these obstructions via examples,
which we expect to be related to the ones on orthogonal structure
groups above.

\medskip
One can, in fact, study the generators $x_9$ and $x_{10}$ a little 
more precisely, by specializing the general disucssion in \cite{St} to our context
to relate to more standard classes, namely the fundamental classes of 
Eilenberg-MacLane spaces. 
We start with 2-Orientations. The spectral sequence of the fibration 
$BO \langle 9 \rangle \to BO \langle 8 \rangle \to K(\Z, 8)$ gives the long exact sequence of 
cohomology groups 
\(
\xymatrix{
\cdots \ar[r] & 
H^9({\rm BO} \langle 9 \rangle; \Z_2) \ar[r]^-\tau & H^{10}(K(\Z, 8); \Z_2) 
\ar[r] &
H^{10}({\rm BO} \langle 8 \rangle; \Z_2) 
\ar[r] 
&
\cdots
}\;.
\label{seq 1}
\)
Here the transgression $\tau$ is given by $x_9 \buildrel{\tau}\over{\mapsto} Sq^2 \iota_8$, where
$\iota_8$ is the fundamental cohomology class of the Eilenberg-MacLane space  $K(\Z, 8)$. 
Next, for 2-Spin structures, 
the spectral sequence of the fibration 
$BO \langle 10 \rangle \to BO \langle 9 \rangle \to K(\Z_2, 9)$ gives the long exact sequence of 
cohomology groups 
\(
\xymatrix{
\cdots \ar[r] & 
H^{10}({\rm BO} \langle 9 \rangle; \Z_2) \ar[r]^-\tau & H^{11}(K(\Z_2, 9); \Z_2) 
\ar[r] &
H^{11}({\rm BO} \langle 9 \rangle; \Z_2) 
\ar[r] 
&
\cdots
}\;.
\label{seq 2}
\)
The transgression $\tau$ is given by $x_{10} \buildrel{\tau}\over{\mapsto} Sq^2 \iota_9$, with
$\iota_9$ the fundamental cohomology class of   $K(\Z_2, 9)$. 

\medskip
Next we consider the cohomology rings the classifying spaces of 2-Orientations and 
2-Spin structures. Let ${\cal A}_2$ be the mod 2 Steenrod algebra and $f_9: {\rm BO}\langle 9 \rangle
=B{\rm Fivebrane} \to  K(\pi_9({\rm BO}), 9) =K(\Z_2, 9)$, 
$f_{10}: {\rm BO}\langle 10 \rangle
={\rm B}2$-Orientation $\to K(\pi_{10}({\rm BO}), 10) =K(\Z_2, 10)$,
and 
$f_{12}: {\rm BO}\langle 12 \rangle= {\rm BO} \langle 11 \rangle 
={\rm B}2$-Spin $\to K(\pi_{12}({\rm BO}), 12) =K(\Z, 12)$,
be the maps realizing the lowest nontrivial cohomology groups,
arising from the lowest nontrivial homotopy groups. 
The induced maps on cohomology are then given by 
$
f_9^*: H^i(K(\Z_2, 9); \Z_2) 
\buildrel{x_9}\over{\longrightarrow}
H^i(B{\rm Fivebrane}; \Z_2)$,
$
f_{10}^*: H^i(K(\Z_2, 10); \Z_2) 
\buildrel{x_{10}}\over{\longrightarrow}
H^i({\rm B2}\mbox{-}{\rm Orientation}; \Z_2)$,
and 
$
f_{12}^*: H^i(K(\Z, 12); \Z_2) 
\buildrel{x_{12}}\over{\longrightarrow}
H^i({\rm B2}\mbox{-}{\rm Spin}; \Z_2)$,
respectively. 
Conditions on the subjectivity of these maps can be deduced from \cite{St2},
and which we will record momentarily. 
We can also deduce from 
the general results of Stong \cite{St2} that the 
cohomology rings with $\Z_2$ coefficients for our spaces are 
\begin{eqnarray}
H^i({\rm BO}\langle 9 \rangle; \Z_2) &\cong& \left({\cal A}_2/ {\cal A}_2 Sq^2 \right)f_9^*(\iota_9)\;,
\nonumber\\
H^i({\rm BO}\langle 10 \rangle; \Z_2) &\cong& \left({\cal A}_2/ {\cal A}_2 Sq^3 \right)f_9^*(\iota_{10})\;,
\nonumber\\
H^i({\rm BO}\langle 12 \rangle; \Z_2) &\cong& \left({\cal A}_2/ {\cal A}_2 Sq^1 + {\cal A}_2 Sq^2 \right)f_9^*(\iota_{12})\;,
\end{eqnarray}
where $\iota_j$ is the fundamental class of the appropriate Eilenberg-MacLane space in degree $j$.

\medskip
We summarize the above discussion with the following
\begin{proposition}
{\bf (i)} The generators $x_9$ and $x_{10}$ are related to the fundamental classes 
$\iota_8$ and $\iota_9$ of $K(\Z, 8)$ and $K(\Z_2, 9)$ via $\tau(x_9)=Sq^2 \iota_8$
and $\tau(x_{10})=Sq^2 \iota_9$, with $\tau$ the transgression in  \eqref{seq 1} and 
\eqref{seq 2}, respectively. 

\noindent {\bf (ii)} The maps $f^*_9$, $f^*_{10}$ and $f^*_{12}$ are surjective for 
$i<18$, $i<20$ and $i< 24$, respectively. 
\end{proposition}

Note that the above inequalities are certainly within the range of dimensions of interest in
string theory and M-theory. 

\medskip
We now go back to the original question on whether $x_9$ and $x_{10}$ have to do with 
Stiefel-Whitney classes. We will deduce from the results of Bahri-Mahowald \cite{BM} and 
Stong \cite{St} that there is no simple direct relation. We consider the covering map 
$p: {\rm BO} \langle \phi(r) \rangle \to {\rm BO}$, where $\phi (r)$ 
is some specific function of $r$, the three relevant values of which are given as
$\phi (3)=8$, $\phi (4)=9$ and $\phi (5)=10$. A result of \cite{St} asserts that 
the covering map $p$ maps $w_i \in H^*({\rm BO}; \Z_2)$ to the generators in
$H^*({\rm BO} \langle \phi (r) \rangle; \Z_2)$ if $i-1$ has a t least $r$ ones in its dyadic 
(binary) expansion, and the remaining classes are mapped to decomposables. 
For our three relevant cases, $i=8, 9$ and 10, we see that the numbers 7, 8, and 9
have numbers of ones in their binary expansions which are certainly smaller than 
$r=4, 5$, and 6, respectively. 

\medskip
Furthermore, a result of \cite{BM} states that the class 
$p^* w_n$ is nonzero in $H^*({\rm BO} \langle \phi (r) \rangle; \Z_2)$ if and only if
a certain Poincar\'e series has a nonzero entry in dimension $n$. For $r=3$ this 
series is given by 
$1 + t^8 + t^{12} + \cdots$, which explicitly contains a $t^8$-term. This is the 
Fivebrane case. Next, for the 2-Orientation case, we have for $r=4$ the 
series 
$
\frac{1}{1- t^{32}}( 1+ t^{16}) (1+ t^{24})(1+ t^{28})(1+ t^{30})(1+ t^{31})$, which does 
{\it not} have a $t^9$ term. Similarly for the case of 2-spun structures, the 
Poincar\'e  series for $r=5$ is likewise sparse and is explicitly seen to no have a $t^{10}$ term. 
Therefore, from both results of \cite{BM} and \cite{St} we have 

\begin{proposition} {\bf (i)}  
The covering maps $p: {\rm BO} \langle 9 \rangle \to {\rm BO}$ and 
 $p: {\rm BO} \langle 10 \rangle \to {\rm BO}$ send the Steifel-Whitney classes
 $w_9 \in H^9({\rm BO}; \Z_2)$
 and $w_{10} \in H^{10}({\rm BO}; \Z_2)$ to decomposables.

\noindent {\bf (ii)} The classes $p^* w_9$ and $p^* w_{10}$ are zero in
$H^9({\rm BO}\langle 9 \rangle; \Z_2)$ and $H^{10}({\rm BO}\langle 10 \rangle; \Z_2)$,
respectively.
\end{proposition}

\section{Ninebrane structures}
\label{sec 9}


In this section we shift from Stiefel-Whitney classes to Pontrjagin classes to 
define our third main structure. 
It is easy to see that $p_1$ is divisible by 2 when we have a String structure. 
In this case, since $w_4$ is the mod  2 reduction of the first Spin characteristic 
class $Q_1=\tfrac{1}{2}p_1$, we have that $w_4=0$ in the presence of a String structure.
From the congruence $p_3=w_6^2$ mod 2 and the fact that $w_6=Sq^2 w_4$ we get
that $p_3$ is even under these conditions. In the four-dimensional case, this was enough to determine 
the obstruction. However, in this twelve-dimensional case, we will see an extra 
 division by $2^2$. Also, as in the case of the Fivebrane structure, 
 there is an extra division by 3 and, additionally, by the next prime 5 for for a Ninebrane
 structure. Another distinction to make is that, while $w_4=0$ does not imply $w_8=0$, 
 having $w_8=0$ does imply $w_{12}=0$. The follows from the Wu formula 
 $Sq^4w_8=w_4 w_8 + w_{12}$, and what distinguishes it from the former is the 
 relatively low power of the Steenrod  square. 

\begin{definition} A {\it Ninebrane structure} 
on a 2-Spin manifold $M$ is a lift $\hat{f}$ of the classifying map $f$ in
the following diagram 
$$
\xymatrix{
&& BO\langle 13 \rangle=B{\rm Ninebrane} 
\ar[d]^\pi  & \\
M \ar@{..>}[rru]^-{\hat{f}} 
\ar[rr]^{~~f} && BO\langle 11\rangle
\ar[r]^{x_{12}}& K(\Z, 12)\;.
}
$$
\end{definition}

The obstruction class 
for lifting the classifying map $f: X \to {\rm BO}\langle 12 \rangle$ to 
$\hat{f}: X \to {\rm BO}\langle 13 \rangle={\rm BNinebrane}$
is obtained by pulling back the universal 
class $x_{12} \in H^{12}({\rm BO}\langle 12\rangle; \Z)\cong H^{12}(BO \langle 11\rangle; \Z)$. 
Thus a manifold $X$
admits a Ninebrane structure if and only if $f^*x_{12}\in H^{12}(X; \Z)$ 
vanishes. This is a fraction of the third Pontrjagin class $p_3$ and 
can be characterized as follows. 
For vector bundles over the sphere $S^{12}$ the best possible result on the
divisibility of the Pontrjagin class $p_3(\xi)$, for $\xi$ a vector bundle of 
rank $12$ over $S^{12}$, is that $p_3(\xi)$ can be any multiple of 
(see \cite{MS} p. 244, \cite{BH2} \cite{Ker})
$
(2\cdot 3-1)! {\rm gcd}(3+1, 2)u$, that is 
$240u$, 
where $u$ is the standard generator of $H^{12}(S^{12})$. 
It follows that the relevant fraction is given by
$1/240$ so that, therefore, we straightforwardly have
\begin{proposition}
The obstruction to a Ninebrane structure is given by $\frac{1}{240}p_3$.
\end{proposition}

\paragraph{Remarks. (i)} 
Note that having simultaneously 
a String structure, a Fivebrane structure, and positive scalar curvature
leads to a Ninebrane structure. This follows from the Lichnerowicz theorem 
and the index theorem: the obstruction to positive scalar curvature is given 
by the $\widehat{A}$-genus, which in dimension 12 is a combination of 
the String obstruction $\tfrac{1}{2}p_1$, the Fivebrane obstruction 
$\tfrac{1}{6}p_2$ and the Ninebrane obstruction $\tfrac{1}{240}p_3$. 
See \cite{Spin} for extensive discussions in the Spin case.  
 
 \vspace{1mm}
\noindent {\bf (ii)} Note that there is a path fibration  
$K(\pi_{12} (B)), 11) \to PK(\pi_{12}(BO), 12) \to K(\pi_{12}(BO), 12)$, 
i.e. $\Omega K(\Z, 12) \to PK(\Z, 12) \to K(\Z, 12)$, which 
 induces the fibration 
 \(
 K(\Z, 11) \longrightarrow {\rm BNinebrane} \longrightarrow {\rm B}2\mbox{-}{\rm Spin}\;.
 \label{9to2} 
 \)

\paragraph{Example.}
Global considerations in M-theory require extension to a 12-dimensional 
bounding Spin manifold $Z^{12}$. Supersymmetry implies the existence of 
a Rarita-Schwinger field, i.e. a spinor-valued one-form, which can be viewed 
as a section of the Spin bundle tensored with the virtual bundle $TZ^{12}_\C -4{\cal O}$,
where the subtraction of $4{\cal O}$ accounts for two Faddeev-Popov ghosts  
as well as the two extra directions in relating to the Spin bundle in ten dimensions \cite{DMW}.
The action can be written in terms of indices of twisted Dirac operators, one of which 
being the Rarita-Scwinger operator \cite{Flux}. 
The Chern character ${\rm ch}(TZ_\C^{12} - 4 {\cal O})$
is given by 
\(
8 + p_1 + \tfrac{1}{12}(p_1^2- 2p_2) + \tfrac{1}{360}(p_1^3 - 3p_1 p_2 + 3p_3)\;.
\)
Equipping our manifold with a String and a Fivebrane structure, i.e. requiring the 
vanishing of $\tfrac{1}{2}p_1$ 
and $\tfrac{1}{6}p_2$, we get in dimension twelve the term $\tfrac{1}{120}p_3$.
This is twice the Ninebrane obstruction $\tfrac{1}{240}p_3$. The situation here is,
in some sense, analogous to  the case of the first Pontrjagin class 
$p_1$ vs. the first Spin characteristic class $Q_1=\tfrac{1}{2}p_1$. Note that, concentrating on 
the prime 2, the Ninebrane obstruction is $\tfrac{1}{8}p_3$, whose mod 2 reduction 
is the Stiefel-Whitney class $w_{12}=0$. 
The similar statement in degree eight is that the 2-adic part of the Fivebrane obstruction which is 
$\tfrac{1}{2}p_2$ admits $w_8$ as its mod 2 reduction. 

\medskip
Note, however, that the above example does not amount to a full anomaly cancellation
requirement, but merely that the Ninebrane obstruction appears in the expressions of part of
the anomaly or effective action. This is then slightly 
weaker that the statements in the String and Fivebrane 
cases, which amounted e.g. to the Green-Schwarz anomaly cancellation condition and its
dual \cite{SSS1}  \cite{SSS2}.

\paragraph{Invariances of the structures under homotopy equivalences.}
We consider whether having one of the three structures defined above 
 is a property that is
invariant under automorphisms. The topological invariance of  
String structures is considered in \cite{KSpin}. 
It is interesting to note that the obstructions $\tfrac{1}{2}p_1$ and 
$\tfrac{1}{6}p_2$ for String and Fivebrane structures,
and  the class $p_3$ mod 120 are homotopy  invariant.
This follows from the results in \cite{Sin}. 
Furthermore, we note that the intersection form on a closed 
Spin 12-dimensional manifold is always even, so we have a further  
division by 2 for $p_3$. Therefore, we have

%
%
%
%

\begin{proposition}
Having a Fivebrane, Ninebrane, 2-Orient, or 2-Spin structure is a homotopy 
invariant property. So if $f: X \to Y$ is a homotopy equivalence  then 
there is such a structure on $X$ if and only if there is the same one on $Y$. 
\end{proposition}

\paragraph{Remark.} In \cite{SSS2} \cite{SSS3}, the identification of anomalies in 
M-theory and string theory with Fivebrane structures required some modification
to account for further congruences. For example, the class $\tfrac{1}{48}p_2$ 
appeared instead of $\tfrac{1}{6}p_2$. That further division by 8 
was accounted for by defining a variant structure, denoted ${\cal{F}}\langle 8 \rangle$.  
Here in the case of Ninebrane structures, the same kind of argument applies and 
we can similarly account for further divisions of $\tfrac{1}{240}p_3$ as needed for 
applications.

\section{The set of lifts}
\label{sec set}

The set of lifts of familiar structures, such as orientations,
Spin structures and String structures is given generally by
a torsor over a cohomology group of one dimension 
less than the dimension of the obstruction. 
This was also shown for Fivebrane structures in \cite{SSS2}. 
Such a characterization 
continues to hold in our case of 
2-${\rm Orientation}={\rm BO}\langle 10\rangle$,
2-${\rm Spin}={\rm BO}\langle 11\rangle$, and 
${\rm Ninebrane}= {\rm BO}\langle 13\rangle$ structures. 
We will describe these lifts in a uniform fashion. 
Let $A$ denote $\Z_2$ for the first and second structures
and $\Z$ for the third structure and let $n=9, 10$, and 12,
respectively, and
$m=n+1$. In the case of the 2-${\rm Spin}$ structure we 
have an automatic further lift one more level to ${\rm BO}\langle 12\rangle$.
We encode all this succinctly in the diagram
\(
\xymatrix{
&
K(A, n-1) 
\ar[d]
\\
&
{\rm BO}\langle m \rangle
\ar[d]
\\
X 
\ar[r]^f
\ar@{..>}[ur]^{\hat{f}}
&
{\rm BO}\langle n \rangle \;,
}
\)
in which the fibrations are induced from the path fibrations,
including the ones in \eqref{path1} and \eqref{path2}. 
The set of structure is given in the three cases by 
a torsor for the cohomology group $H^{n-1}(X; A)$. 
Therefore, with an equivalence relation on each of the sets given by 
homotopy of sections, we have

\begin{proposition}
(i) The set of 2-Orientation structures on a given Fivebrane structure is given by 
a torsor for the group $H^8(X;\Z_2)$.

\noindent (ii) The set of 2-Spin structures on a given 2-Orientation structure  is given by 
a torsor for the group $H^9(X;\Z_2)$.

\noindent (iii) The set of Ninebrane structures on a given 2-Spin structure  is given by 
a torsor for the group $H^{11}(X;\Z)$.
\end{proposition}

\paragraph{Remarks.} We note the following:

\noindent {\bf 1.} On manifolds $Y^{11}$ of dimension eleven, i.e. as in M-theory,
the Ninebrane obstruction vanishes identically by dimension reasons. However, it is 
still interesting to consider Ninebrane structures on $Y^{11}$ as those are 
parametrized by the group $H^{11}(Y^{11}; \Z)$. This is analogous to the case of 
String structures on 3-dimensional manifolds $M^3$, where these structures are
parametrized by $H^3(M^3; \Z)$, corresponding to a gerbe on the worldvolume and 
is captured by the volume; see \cite{tcu} for a characterization and application  
to the M2-brane. The second case is having a Fivebrane structure on the worldvolume 
$M^6$ of the M5-brane. Again this is automatic, but the structures are interestingly 
enumerated by
the 5-gerbe on the worldvolume. See \cite{SSS3} \cite{FSS1} \cite{inv} 
for detailed accounts.

\noindent {\bf 2.} In lower degrees, the lift to a 
${\rm BO}\langle n+1 \rangle$-structure does not 
depend on the choice of a ${\rm BO}\langle n\rangle$-structure.
This is the case for $n=2$, where the existence of a Spin structure does not depend on 
choice of orientation from which to lift,  because of homotopy invariance 
of the second Steifel-Whitney class $w_2$. 
The case  $n=4$ is similar, where a 
lift to a String structure does not depend on a choice of 
underlying Spin structure from which to lift, which can be shown via obstruction theory
(see \cite{CCV}).
However, in higher degrees this changes -- see 
The Manifold Atlas Project \cite{Atlas}, which we follow in the ensuing discussion.
 Starting with 
Fivebrane structures and going up, one has dependence of the
higher structure on the choice of lower structures. 
That is, among the set of String structures there might exist
one which does not lift to a Fivebrane structure. 
Let us illustrate the statement in this degree, and the next degrees which we consider in this
article will follow analogously. Consider two ${\rm String}$ structures 
given by two classifying maps $f, g: X \to {\rm BString}$
for which the composition $\hat{f}, \hat{g}: 
X  \to {\rm BFivebrane} \buildrel{\pi}\over{\to} {\rm BString}$ are homotopic. 
Then the two maps $f,g$ differ by a map $h: X \to K(\Z, 7)$ to
the homotopy fiber of $\pi$. We have the diagram
\(
\xymatrix{
&&&& 
K(\Z, 7)
\ar[d]
\\
&&
~~~~~~~~~{\rm BO}\langle 9 \rangle
\times K(\Z, 7) 
\ar[r]^-{id \times i}
&
{\rm BO}\langle 9 \rangle\times {\rm BO}\langle 9 \rangle
\ar[r]^-{m}
&
{\rm BO}\langle 9 \rangle
\ar[d]^\pi
\\
X 
\ar@/^5.3pc/[urrrr]^{\hat{f}}
\ar[urr]_{\hspace{1mm}\hat{g}, h}
\ar[rrrr]^{f, g}
&&&&
{\rm BString}\;,
}
\)
where $m$ is the $H$-space multiplication. 
 The induced maps on the corresponding cohomology groups are captured by the 
 sequence 
 \footnote{We do not claim that we know the structure of the cohomology ring
 $H^*({\rm BO}\langle 9 \rangle; \Z_2)$, but it is
 enough for us to know the first generator and that there is an H-space structure. 
 For rational coefficients, this is 
 studied in \cite{SSS2}.}
 \begin{eqnarray}
 \hspace{-.5cm}
 H^*(X; \Z) 
 \to
 H^*({\rm BO}\langle 9 \rangle; \Z_2) \times K(\Z, 7) 
 \to
 H^*( {\rm BO}\langle 9 \rangle; \Z_2) \otimes H^*({\rm BO}\langle 9 \rangle; \Z_2)&&
 \nonumber\\
 \hspace{5cm} \subseteq
H^*({\rm BO}\langle 9 \rangle \times {\rm BO}\langle 9 \rangle; \Z_2)&
 \longrightarrow&
 H^*({\rm BO}\langle 9 \rangle; \Z_2)
 \nonumber
 \end{eqnarray}
 \(
 \hspace{-15mm}
 \xymatrix{
 ~\hat{f}^*x_9=\hat{g}^*x_9 + h^* i^*x_9
  \ar@{<-|}[r]
 &
 x_9\otimes 1 + 1 \otimes i^*x_9
  \ar@{<-|}[rr]
 &&
 x_9\otimes 1 + 1 \otimes x_9
 \ar@{<-|}[r]
 &
 x_9:=k\;.
 }
 \nonumber
 \)
%

\medskip
Now we take $X=K(\Z, 7)$ and $h$ the identity. Then in this case,
we investigate whether the pullback of the characteristic class 
$k=x_9 \in H^9({\rm BFivebrane};\Z_2)$ under the two maps 
$\hat{f}$ and $\hat{g}$ agree. Since 
$\hat{f}^*x_9=\hat{g}^*x_9 + h^* i^*x_9$, this question reduces to whether 
or not the pullback of $x_9$ under 
$i: K(\Z, 7) \to {\rm BFivebrane}$ is zero.
Using \cite{St}, we have that $i^*(x_9)=Sq^2 \iota_7$, where 
$Sq^2$ is the Steenrod square of degree two and $\iota_7$ is the 
fundamental class of $K(\Z, 7)$. 
As this pullback is not zero, we have that the two classes
$\hat{f}^*x_9$ and $\hat{g}^*x_9$ do not agree. 
In particular, if one of them is zero then the other is not zero.
Consequently, it can be arranged that one Fivebrane structure can lift 
to  a ${\rm BO}\langle 10 \rangle$-structure while the other
cannot. 

\medskip
The discussion for lifting further from ${\rm BO}\langle 10 \rangle$ to 
${\rm BO} \langle 11\rangle$ one encounters the pullback of 
a degree ten class $x_{10}$ via the map $i: K(\Z_2, 8) \to BO\langle 10\rangle$.
Using \cite{St}, such a map is given by $i^*(x_{10})=Sq^2 \iota_8$, where
$\iota_8$ is the fundamental class of $K(\Z_2, 8)$. 
Therefore, the obstruction classes corresponding to two homotopic maps 
can be different and, again, it can be arranged that one is trivial while the 
other is not, leading to a lift for only one of them.
The story for the lift to Ninebrane is similar. 
Furthermore, the discussion can be extended similarly 
for $X$ other than an Eilenberg-MacLane space. Using the 
discussion in Chap. XI of \cite{St2} 
the corresponding classes
$x_9$, $x_{10}$ and $x_{12}$ 
belong to (cosets of) $Sq^2 \rho_2 H^7(X;\Z)$, 
 $Sq^2  H^8(X;\Z_2)$, and $Sq^3 H^9(X;\Z_2)$, respectively.

\section{Twisted structures}
\label{sec twist} 
In this section we define twisted versions of the structures defined above,
using the approach in \cite{Wang} \cite{SSS3} \cite{tw1} \cite{tw2}.

\begin{definition}(Twisted 2-orientation)
A twisted 2-orientation on a submanifold (a brane) $M$ embedded in spacetime $X$ is 
a homotopy in the following diagram, where $f$ is the classifying map
for Fivebrane bundles and 
 $\alpha_9$ is a cocycle of degree 9
$$
\xymatrix{
M \ar[rr]^-{f}_>{\ }="s"
\ar[d]^i
&&
{\rm BO\langle 9\rangle}
\ar[d]^{x_9}
\\
X \ar[rr]^{\alpha_9}^<{\ }="t" 
&&
K(\Z_2, 9)\;.
 \ar@{=>}_\eta "s"; "t"
}
$$
\end{definition}

\paragraph{Remark.} The obstruction to having a twisted 2-orientation on $M$ is given by 
\(
f^*x_9 + i^* \alpha_9=0\;.
\)

\begin{definition} (Twisted 2-Spin structure)
A twisted 2-Spin structure on a submanifold (a brane) $M$ embedded in spacetime $X$ is 
a homotopy in the following diagram, where $f$ is the classifying map of 2-oriented bundles 
and  $\alpha_{10}$ is a cocycle of degree 10
$$
\xymatrix{
M \ar[rr]^-{f}_>{\ }="s"
\ar[d]^i
&&
{\rm BO}\langle 10\rangle
\ar[d]^{x_{10}}
\\
X \ar[rr]^{\alpha_{10}}^<{\ }="t"  
&&
K(\Z_2, 10)\;.
 \ar@{=>}_\eta "s"; "t"
}
$$
\end{definition}
\paragraph{Remark.} The obstruction to having a twisted 2-Spin structure on $M$ is given by 
\(
f^*x_{10} + i^* \alpha_{10}=0\;.
\)
It would be interesting to provide examples of twisted 2-Spin structures and 
twisted 2-Orientation structures, along the lines of \cite{tw1} \cite{tw2}. 


\begin{definition}
(Twisted Ninebrane structure)
A twisted Ninebrane structure is defined by the following diagram 
$$
\xymatrix{
M \ar[rr]^f_>{\ }="s" 
\ar[d]^i
&&
{\rm BO}\langle 12 \rangle
\ar[d]^{\tfrac{1}{240}p_3}
\\
X \ar[rr]^{\alpha_{12}}^<{\ }="t"   
&&
K(\Z, 12)\;.
 \ar@{=>}_\eta "s"; "t"
}
$$
\end{definition}

\paragraph{Remark.}
The obstruction to having a twisted Ninebrane structure is given by 
\(
f^*(\tfrac{1}{240}p_3) + i^* \alpha_{12}=0\;.
\)

\paragraph{Example: Twisted Ninebrane structures via embeddings.}
We consider a brane embedded in spacetime via $i: M \hookrightarrow Z$.
Then the $\widehat{A}$-genera of $M$ and $Z$ can be related via 
a Riemann-Roch formula. In the simplest case where this embedding is 
a homotopy equivalence, one has that $\widehat{A}(Z)/\widehat{A}(M)$
is an element in the real Chern character ${\rm chO}(Z)$, that is a Pontrjagin 
class of some orthogonal bundle \cite{AH}. Considering degree four components
gives $\tfrac{1}{2}p_1(M)=\tfrac{1}{2}p_1(Z) + 12 p_1(E)$, where $E$ is an 
orthogonal bundle on $Z$. Then a String structure on $M$ leads to a 
twisted String structure on $Z$ as $12p_1(E) \in \Z$. This can 
also be reversed; by writing  $\tfrac{1}{2}p_1(M)- 12 p_1(E)=\tfrac{1}{2}p_1(Z)$,
a String structure on $Z$ amounts to a twisted String structure on $M$.
In the presence of a Spin structure, the statement can be improved to
a further divisibility by two due to the congruence $p_1^2=w_2^2$ mod 2. 
By the above general formula of Atiyah-Hirzebruch 
\cite{AH} we can deduce similar statements in 
higher degree cases (and following the approach of \cite{tcu2} with sufficiently high dimensions): 

\noindent {\bf 1.} {\it Degree eight:} In the presence of a String structure 
on both $M$ and 
$Z$, the degree eight components give 
\(
\tfrac{1}{6}p_2(M) = \tfrac{1}{6}p_2(Z) + 240 p_2(E)\;.
\)
We view this as an example of a twisted
Fivebrane structure on $Z$ determined by a Fivebrane 
structure on $M$ and vice versa.

\noindent {\bf 2.} {\it Degree twelve:} Assuming String structures 
$\tfrac{1}{2}p_1(M)=0=\tfrac{1}{2}p_1(Z)$
and Fivebrane structures $\tfrac{1}{6}p_2(M)=0=\tfrac{1}{6}p_2(Z)$ 
on both the brane and spacetime, 
we have in degree twelve
\(
\tfrac{1}{240}p_3(M) = \tfrac{1}{240}p_3(Z) + 252 p_3(E)\;.
\)
Therefore, upon setting each side to zero, 
this gives an 
equivalence between a Ninebrane structure on $M$ and a twisted
Ninebrane structure on $Z$, and vice versa.

\paragraph{Example: The $E_8$ index in M-theory.}
The index of the Dirac operator coupled to an $E_8$ bundle in M-theory on a Spin manifold $Y^{11}$
lifted to twelve dimensions 
is given by 
\(
\tfrac{1}{2}I_{E_8}=\tfrac{1}{6}G_4 \cup G_4 \cup G_4 
- \tfrac{1}{48}(p_2 - (\tfrac{1}{2}p_1)^2)\cup G_4 
- \tfrac{31}{15120} p_3 + \tfrac{13}{30240}p_1p_2
- \tfrac{1}{15120}p_1^3\;.
\label{E8}
\)
Assuming a String structure, the C-field quantization condition \cite{Flux} 
$G_4 +\tfrac{1}{4}p_1=a \in H^4(Y^{11}; \Z)$ reduces to $G_4=a$, the characteristic class of the $E_8$ 
bundle. If we further assume a Fivebrane structure, i.e. $\tfrac{1}{6}p_2=0$, then 
expression \eqref{E8} reduces to 
\(
\tfrac{1}{2}I_{E_8}=\alpha \cdot \tfrac{1}{240}p_3 - \tfrac{1}{6}a\cup a \cup a\;,
\)
where $\alpha=\tfrac{31}{63}$. 
For any value of $\alpha$, the second term can viewed as a rational twist for 
a rational Ninebrane structure. 
However, when $1/\alpha$ times the last term is integral, 
this latter term serves as an integral twist
for the first term, which is an 
obstruction to the would-be Ninebrane structure. 
Therefore, the triviality of the Dirac index for $E_8$ bundles with classes
$a=186n$ for $n\in \Z$, in M-theory on a
twelve-dimensional Fivebrane manifold $Z^{12}$ is equivalent to a twisted
Ninebrane structure  on $Z^{12}$, with the twist given by the cubic term in 
the $E_8$ characteristic class. Note that this kind of twist is composite and is a 
cubic analog in degree twelve of the composite quadratic twists giving 
rise to a ${\rm String}^c$ structure in degree four
\cite{tcu2} and 
a ${\rm Fivebrane}^{K(\Z, 4)}$ structure in degree eight \cite{tw1}.

\medskip
Again this example highlights the fact that due to the relative high dimension 
of the Ninebrane obstruction relative to the dimensions of the applications
considered, the situation is not as optimal as one had in the cases of 
lower obstructions, namely of the String and the Fivebrane, in \cite{SSS2} \cite{SSS3}. 
Furthermore, the anomalies encountered should necessarily not be of the 
usual Green-Schwarz type, since these are always of a factorized form: a
product of a degree four piece and a degree eight piece. 
However, we will see in Sec. \ref{sec dif} that there is a natural explanation of a new phenomenon, 
namely the existence of a Chern-Simons term and a
top form in M-theory that lends itself to a natural 
description in terms of Ninebrane structures.

\section{Structures not directly defined via the Whitehead tower}
\label{sec variant}

We know that one can define structures arising from vanishing of (multiples of) 
higher obstructions, without the lower obstructions necessarily vanishing. 
Examples of such are abundant: A Pin structure requires the vanishing of 
the would-be Spin obstruction $w_2$ without necessarily having the orientation 
obstruction $w_1$ vanish. Also, we can have the first Pontrjagin class
$p_1$ vanishing without the lower obstruction, the Spin obstruction $w_2$,
being zero. Such a structure is called a $p_1$-structure and is important in 
Chern-Simons theory and low-dimensional topology. See \cite{tw1} \cite{tw2} \cite{inv} 
for many examples of structures defined via this general phenomenon. 

\medskip
 Let $X={\rm BO}\langle p_i \rangle$ be the homotopy fiber of the map 
$p_i: BO \to K(\Z, 4i)$ corresponding to 
the first Pontrjagin class of the universal stable bundle $\gamma$ 
over the classifying space $BO$. Let $\gamma_X$ be the pullback of 
$\gamma$ over $X$. 
\begin{definition} 
A $p_i$-structure on a submanifold (a brane) $M$ is a fiber
map from the stable tangent bundle $TM$ of $M$ to $\gamma_X$. 
That is, there is the following lifting diagram
$$
\xymatrix{
&& X={\rm BO}\langle p_i \rangle
\ar[d]
&&
\\
M
\ar@{..>}[urr]
\ar[rr]
&& 
{\rm BO}
\ar[r]^-{p_i}
&
K(\Z, 4i)\;.
}
$$
\end{definition} 
The Spin/String version of this 
 construction is explained in our context in 
\cite{tcu}. 
So a $p_2$-structure is defined when $p_2=0$ but $p_1\neq 0$,
and a $p_3$-structure is defined when $p_3=0$ while $p_i\neq 0$, $i=1,2$.

\paragraph{Remark.} 
We can similarly consider structures defined by Stiefel-Whitney classes
and Wu classes, as in \cite{tw2}. 
For instance, we define a {\it 2-Pin structure} via $x_{10}=0$ but $x_9\neq 0$.

%
%
%
%
%

\medskip
We now consider twists of the above structures, generalizing the definition in \cite{inv} 
from degree four to other (higher) degrees. 

\begin{definition}
An $\alpha$-twisted $p_i$-structure on a submanifold (a brane) $\iota: M \to Y$ 
with a Riemannian structure classifying map $f: M \to BO$, is a $4i$-cocycle
$\alpha: Y \to K(\Z,4i)$ 
and a homotopy $\eta$ in the diagram 
$$
    \raisebox{20pt}{
    \xymatrix{
       M
       \ar[rr]^-{f}_>{\ }="s"
       \ar[d]_\iota
       &&
       B \mathrm{O}(n)
       \ar[d]^-{p_i}
       \\
       Y
       \ar[rr]_{\alpha_{4i}}^<{\ }="t"
       &&
       K(\mathbb{Z},4i)\;.
       \ar@{=>}^\eta "s"; "t"
    }
    }
  $$
  \label{def}
\end{definition}

\paragraph{Remarks. (i)} 
The obstruction is then $p_i(M) + [\alpha_{4i}]=0\in H^{4i}(M; \Z)$. 
As in the case of twisted String, Fivebrane, or Ninebrane structures, 
the set of such structures will be a torsor for $H^{4i-1}(M; \Z)$. 

%
%
%
%
%

\noindent {\bf (ii)} Similarly we can define a {\it twisted 2-Pin structure} 
and other variants as the case of those given by Stiefel-Whitney classes and 
Wu classes.

\section{The (twisted) groups} 
\label{sec group}

We have defined the structures directly via classifying spaces in previous sections.
A natural question then is whether and how to describe the corresponding groups
(in the homotopy sense).  Here we build the groups as the deloopings of the classifying 
spaces as in previous cases \cite{SSS3}. The general machinery there allows similarly 
that we define new groups here as follows. 

\begin{definition} 
The homotopy fibers of the structures $B2$-Orient, $B2$-Spin, and $B$Ninbrane define 
the groups 2-Oient, 2-Spin, and Ninebrane, respectively. 
\end{definition}

\paragraph{Remarks. 1.} Working not necessarily in the stable range, we have the 
groups 2-${\rm Oient}(n)$, 2-${\rm Spin}(n)$, and ${\rm Ninebrane}(n)$. In the 
notation for connected covers with conventions as in \cite{SSS3}, 
these groups are ${\rm O}\langle 8\rangle(n)$, ${\rm O}\langle 9 \rangle(n)$ 
and ${\rm O}\langle 11 \rangle(n)$.

\noindent {\bf 2.} The group 2-Orient$(n)$ is the $\Z_2$ double cover (in the homotopy sense, and with 
a mod 8 shift from the classical notion) of the group Fivebrane$(n)$. We have 
$\pi_8({\rm Fivebrane}(n))\cong \Z_2$ while $\pi_8(2$-${\rm Orient}(n))= 0$.

\noindent {\bf 3.} The group 2-Spin$(n)$ is the $\Z_2$ double cover (also in the above sense) 
of the group 2-Orient$(n)$. We have 
$\pi_9(2$-${\rm Orient}(n))\cong \Z_2$ while $\pi_9(2$-${\rm Spin}(n))= 0$.

\noindent {\bf 4.} For any of the above groups $G$ we have $\pi_{10}(G)=0$.
This is a mod 8 shift of the classical fact that $\pi_2(G)=0$ for any Lie group 
(and hence also for its connected covers).

\medskip
Similarly, we can define groups (again in the homotopy sense) as the homotopy 
fibers of the corresponding twisted structures, again as in \cite{SSS3} (see also
\cite{tcu2}). 

\begin{definition} 
The twisted groups $G^c=$ ${\rm O}\langle 8\rangle^c(n)$, ${\rm O}\langle 9 \rangle^c(n)$ 
and ${\rm Ninebrane}^c(n)$
are the homotopy fibers of the corresponding twisted structures $BG^c$. 
\end{definition}

\paragraph{Remark.} As in the cases of String($n$) and Fivebrane($n$),
the Whitehead tower construction allows us to
 describe the group Ninebrane$(n)$ via a fibration with an 
Eilenberg-MacLane space as a fiber
\(
K(\Z, 10) \longrightarrow {\rm Ninebrane}(n) \longrightarrow {2{\mbox{-}{\rm Spin}}(n)}\;,
\) 
obtained by looping the fibration \eqref{9to2}. 

\medskip
\noindent It would be interesting to find explicit geometric/categorical models for the above groups.

 \section{Differential refinement}
 \label{sec dif}
 
 It is desirable for physics to have differential versions of the topological 
 structures that arise. As in the case of String and Fivebrane 
 structures \cite{FStS} \cite{SSS3}, one can consider differential refinements of the 
 higher ${\rm BO}\langle m \rangle$-structures to higher stacks.
 Via the formulation in \cite{FStS} \cite{Sch} we refine the classifying spaces 
 $BG$ as topological spaces to $\mathbf{B}G$ as stacks.
 This also requires refining Eilenberg-MacLane spaces 
 $K(\Z, n)=B^{n-1}U(1)$ to stacks
 $\mathbf{B}^{n-1}U(1)$. Consequently, we have

 \begin{proposition} 
 The above structures  refine to moduli stacks described in the diagram
 \(
\xymatrix{
  \mathbf{B}\mathrm{Ninebrane} 
    \ar[d]_{\mathbf{NinebraneStruc}}
	&& 
     \\
  \mathbf{B}O\langle 11\rangle \ar[rr]^{\tfrac{1}{240}{\bf p}_3} 
     \ar[d]_{\mathbf{2\mbox{-}SpinStr}}
     &&
     \mathbf{B}^{11}U(1)
     \\
   \mathbf{B}O\langle 10 \rangle \ar[rr]^{{\bf x}_{10}}
    \ar[d]_{\mathbf{2\mbox{-}OrientStruc}}
     &&
     \mathbf{B}^{10}\Z_2
     \\
    \mathbf{B}\mathrm{Fivebrane} \ar[rr]^{{\bf x}_9} 
    \ar[d]_{\mathbf{FivebraneStruc}}
	&& \mathbf{B}^9 \Z_2
    \\
 \mathbf{B} \mathrm{String} \ar[rr]^{\tfrac{1}{6}{\bf p}_2} 
    && \mathbf{B}^7U(1)\;.
     }
\)
\end{proposition} 


\paragraph{Remarks.}
{\bf 1.} Strictly speaking, the construction used in \cite{FStS} to 
Lie integrate the first two invariant polynomials of $\mathfrak{so}(n)$ 
to the smooth $\tfrac{1}{2} {\bf p}_1$ and the
smooth $\tfrac{1}{6} {\bf p}_2$   would yield for the third invariant polynomial some
multiple of ${\bf p}_3$  whose homotopy fiber is the result of
killing $\pi_{11} = \Z$ in Fivebrane,  but leaving the $\pi_8 = \pi_9 = \Z_2$ alone.
The smooth $\tfrac{1}{240} {\bf p}_3$ as displayed above exists,
but this does not follow from the construction in
\cite{FStS}.  That construction only kills
cocycles at the level of $L_\infty$ algebras  and then integrates that
up to smooth higher stacks, but so it cannot kill torsion groups.
We thank Urs Schreiber for illuminating discussions on these matters
(see also \cite{Sch}).

\noindent {\bf 2.}We can also provide further refinement
 to $\mathbf{B}^nU(1)_{\rm conn}$ by including connections,
giving a diagram as above but with the moduli stacks of $n$-bundles with 
connections, using the machinery developed in \cite{FStS} \cite{Sch}. 
The corresponding diagram will be one replacing the above, with 
$\mathbf{B}^nA_{\rm conn}$ replacing  $\mathbf{B}^nA$
and the refined classes $\hat{\bf c}$ replacing the classes ${\bf c}$
via \cite{FStS} \cite{SSS3} \cite{Sch}.


\paragraph{Trivialization of the Ninebrane obstruction class.}
Recall that in the case of String and Fivebrane structures we had trivializations 
of the corresponding forms 
given by a degree three class $H_3$ and a degree seven class $H_7$, respectively, 
in essentially the following form 
\(
dH_3=\tfrac{1}{2}p_1(A)\;, \qquad \qquad dH_7=\tfrac{1}{6}p_2(A)\;.
\)
Furthermore, such expressions arise in physical settings, e.g. essentially in the 
Green-Schwarz anomaly cancellation and its dual, as explained in \cite{SSS2} \cite{SSS3}. 
It makes sense to consider for the cohomology class obstructing the Ninebrane 
structure a trivialization at the level of differential form representatives given by 
\(
dH_{11}=\tfrac{1}{240}p_3(A)\;,
\label{9}
\)
for some 11-form $H_{11}$. We investigate whether the form $H_{11}$ 
has some physical interpretation. Note that because of the relatively low dimensions
in M-theory and string theory in comparison to our increasing levels in the Whitehead tower, such
an interpretation becomes harder to get. However, we propose  
a conjectural relation. Recently, existence of top forms was discovered in string theory
(see \cite{BHHOR} and references therein); these are ``potentials" rather than field strengths, i.e.
are higher connections rather than higher curvatures. So such 
a top form $H_{11}$ in M-theory can be taken to satisfy
\(
dH_{11}=-\tfrac{1}{2}G_4 \wedge G_8 + \cdots.
\label{H11 second}
\)
Not much is known about the dynamics or the geometry associated with this form. 
We propose that $H_{11}$ in \eqref{H11 second}  is the trivialization 
of the Ninebrane form \eqref{9}, i.e. the two expressions are compatible in the 
sense that the second admits a correction term by the first. This is analogous in degree 
twelve to the correction of the equations of motion of the C-field by the one-loop 
polynomial $I_8$ in degree eight. We hope that more investigations are made on such forms so that the 
above proposal can be verified. 
We do, however, provide another possible interpretation. 
 Chern-Simons terms $CS_{11}$ of degree 11 appear in the M-theory action when formulated
 via the signature (which is equivalent to formulation via Dirac operators) in \cite{sig}. We have
the relation to the Ninebrane structures and to $p_3$-structures as 
$p_3(A) \sim dCS_{11}(A)$. 

\paragraph{Secondary 9-brane structures.} 
Note that while the 12-class $\sim p3$ necessarily vanishes in 11 dimensions, 
it is noteworthy that the differential refinement to $\hat p_3$ is a 
differential cohomology class in degree 12, it is a {\it secondary
class/invariant} which need not vanish even if its underlying (topological) 12-class
vanishes. This is of course just the statement that there may be a
non-trivial connection 11-form, even if its curvature vanishes.
So while in 11 dimensions any bare Fivebrane structure always has a lift to a
Ninebrane structure, if one considers differential Fivebrane
structures (i.e. maps to ${\bf B} {\rm Fivebrane}_{\rm conn}$) 
then there is a actual condition
to lift to ${\bf B}{\rm Ninebrane}_{\rm conn}$, namely that not just the curvature
12-class but also the connection 11 form itself vanishes.
As indicated above, that 11-form is just the Lagrangian for the 11-dimensional
Chern-Simons term.

\paragraph{Relation to the M-algebra and the M9-brane.} 
In the discussion of the algebra corresponding to eleven-dimensional supergravity, and its
associated cohomology, it was found in \cite{WZW} in the super geometric setting that 
there exists a spacetime-filling brane in M-theory. 
In the case of the M9-brane 
 the relation to generalized Wess-Zumino-Witten (WZW) models and 
 generalized Chern-Simons (CS) theories  is the last
 column in the following table, which completes the first two cases
 studied extensively in \cite{CJMSW} \cite{AJ} \cite{Wal} \cite{tcu} and 
 \cite{tcu} \cite{FSS1} \cite{FSS2} \cite{FSS3} \cite{WZW}, respectively.

\begin{center}
\begin{tabular}{|c||c|c|c|}
\hline
 & String  &      Fivebrane   &     Ninebrane \\
\hline  \hline
Worldvolume &  $\Sigma_2$ & $\Sigma_6$  & $\Sigma_{10}$\\
\hline
WZW & ${\rm WZW}_2$ & ${\rm WZW}_6$ & ${\rm WZW}_{10}$\\ 
\hline
Handlebody  & $M^3$ & $M^7$ & $M^{11}$\\
\hline
Chern-Simons & $CS_3$ & $CS_7$ & $CS_{11}$\\
\hline
Structure & String or $p_1$ & Fivebrane or $p_2$ & Ninebrane or $p_3$\\ 
\hline
\end{tabular}
\end{center}


\medskip
The ${\rm WZW}_{10}$ theory was studied in \cite{gerbes} \cite{loop} \cite{framed} 
\cite{WZW}. 
Both of the theories (${\rm WZW}_{10}$ and $CS_{11}$) associated to the ninebrane 
can be refined to the level of moduli stacks of higher bundles with higher connections,
as lower degrees \cite{FStS} \cite{FSS1} \cite{FSS2} \cite{FSS3} \cite{WZW} \cite{Sch}. 
Analogously to the degree four and degree eight cases, i.e. for String and 
Fivebrane structures, respectively, from the above works,
%
the geometric and topological ingredients associated to the ninebrane are described by the 
following diagram (cf. \cite{Sch})
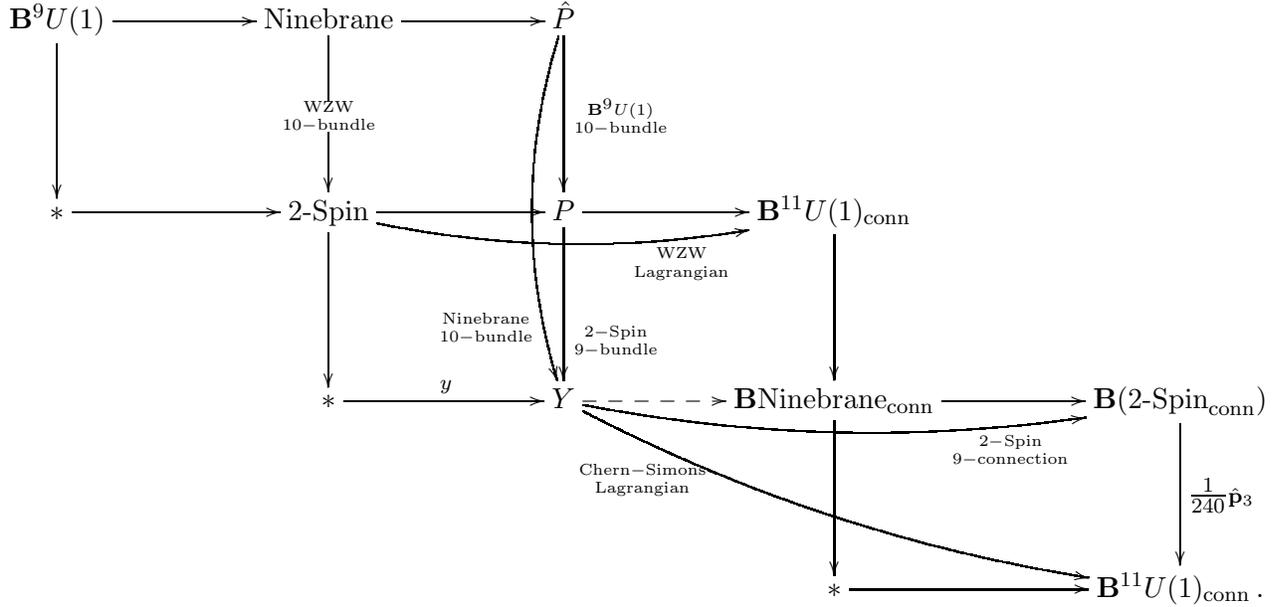
$$
  \xymatrix{
    \mathbf{B}^9 U(1) \ar[rr] \ar[dd] && \mathrm{Ninebrane} \ar[rr] \ar[dd]|{\mathrm{WZW}\atop \mathrm{10{-}bundle}} && 
	 \hat P 
	 \ar[dd]^{\mathbf{B}^9 U(1) \atop \mathrm{10{-}bundle}}
	 \ar@/_1pc/[dddd]_>>>>>>>>{\mathrm{Ninebrane} \atop \mathrm{10{-}bundle}}
	&& 
    \\
	\\
    {*} \ar[rr] && \mathrm{2\mbox{-}Spin} \ar[rr] \ar[dd] 
	\ar@/_1pc/[rrrr]_>>>>>{{\mathrm{WZW} \atop \mathrm{Lagrangian}}}
	&& P \ar[rr] \ar[dd]^>>>>>>{\mathrm{2-Spin}\atop \mathrm{9{-}bundle}} 
	&& \mathbf{B}^{11} U(1)_{\mathrm{conn}}  \ar[dd] 
	  &&  &&
    \\
	\\
    && {*} \ar[rr]^y && Y \ar@/_1pc/[rrrr]_>>>>>>{{\mathrm{2{-}Spin} \atop \mathrm{9{-}connection}}} \ar@{-->}[rr] 
	\ar@/_1pc/[rrrrdd]_<<<<<<<<<{{\mathrm{Chern{-}Simons} \atop \mathrm{Lagrangian}}}
	&& \mathbf{B} \mathrm{Ninebrane}_{\mathrm{conn}} \ar[rr] \ar[dd] && 
	  \mathbf{B} (\mathrm{2\mbox{-}Spin}_{\mathrm{conn}}) \ar[dd]^{\tfrac{1}{240}\hat {\mathbf{p}}_3} &&
	\\
	\\
	&& && && {*} \ar[rr] && \mathbf{B}^{11} U(1)_{\mathrm{conn}}
  \,.
  }
$$
Note that this diagram is such that all squares
(and hence all composite rectangles) are homotopy cartesian (i.e. are
homotopy pullback squares) in the $\infty$-topos of smooth
$\infty$-groupoids (i.e. $\infty$-stacks over smooth manifolds). This
is a much stronger statement than just that the diagram exists, as 
each item in the top-left of a square is in fact
uniquely characterized (up to equivalence) as completing that square
to a homotopy pullback square. Similarly, for the case of $p_3$-structures
$$
  
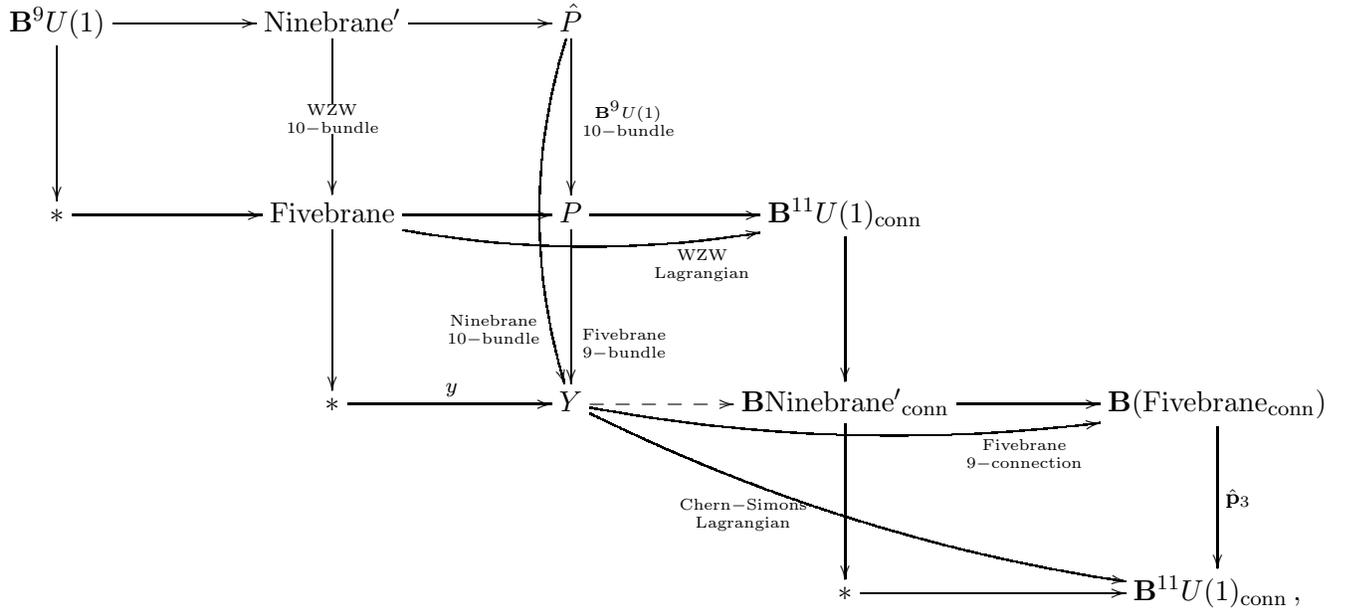
\xymatrix{
    \mathbf{B}^9 U(1) \ar[rr] \ar[dd] && \mathrm{Ninebrane'} \ar[rr] \ar[dd]|{\mathrm{WZW}\atop \mathrm{10{-}bundle}} && 
	 \hat P 
	 \ar[dd]^{\mathbf{B}^9 U(1) \atop \mathrm{10{-}bundle}}
	 \ar@/_1pc/[dddd]_>>>>>>>>{\mathrm{Ninebrane} \atop \mathrm{10{-}bundle}}
	&& 
    \\
	\\
    {*} \ar[rr] && \mathrm{Fivebrane} \ar[rr] \ar[dd] 
	\ar@/_1pc/[rrrr]_>>>>>{{\mathrm{WZW} \atop \mathrm{Lagrangian}}}
	&& P \ar[rr] \ar[dd]^>>>>>>{\mathrm{Fivebrane}\atop \mathrm{9{-}bundle}} 
	&& \mathbf{B}^{11} U(1)_{\mathrm{conn}}  \ar[dd] 
	  &&  &&
    \\
	\\
    && {*} \ar[rr]^y && Y \ar@/_1pc/[rrrr]_>>>>>>{{\mathrm{Fivebrane} \atop \mathrm{9{-}connection}}} \ar@{-->}[rr] 
	\ar@/_1pc/[rrrrdd]_<<<<<<<<<{{\mathrm{Chern{-}Simons} \atop \mathrm{Lagrangian}}}
	&& \mathbf{B} \mathrm{Ninebrane'}_{\mathrm{conn}} \ar[rr] \ar[dd] && 
	  \mathbf{B} (\mathrm{Fivebrane}_{\mathrm{conn}}) \ar[dd]^{\hat {\mathbf{p}}_3} &&
	\\
	\\
	&& && && {*} \ar[rr] && \mathbf{B}^{11} U(1)_{\mathrm{conn}}
  \,,
  }
$$
with Ninebrane$'$ referring to a $p_3$-structure. The main difference between the two diagrams above
is that the second does not involve killing the two $\Z_2$'s in degrees 9 and 10. 
Note that the latter diagram, unlike the former, 
is a corollary of the theorem
in \cite{FStS} -- see the Remarks at the beginning of this section.

  \bigskip\bigskip
\noindent
{\bf \large Acknowledgements}\\
 

\noindent The author thanks Domenico Fiorenza, Urs Schreiber, and Jim Stasheff
for enjoyable collaborations on projects on which this one builds and for useful discussions.  
The author is grateful to Martin Olbermann for pointing out and emphasizing subtleties in 
identifying the generators and to Urs Schreiber for valuable comments on the manuscript. 
The author also thanks the anonymous referee for useful suggestions to improve the 
paper. 
This research is supported by NSF Grant PHY-1102218.


\end{document}